\documentclass[reprint,prb,superscriptaddress,dvipdfmx,amsmath]{revtex4-1}
\usepackage{graphicx,color}

\newcommand{\sinc}{\text{sinc}}
\renewcommand{\i}{\text{i}}
\newcommand{\s}{\text{s}}
\newcommand{\vac}{\text{vac}}
\newcommand{\pdc}{\text{twin}}
\newcommand{\en}{\text{e}}
\newcommand{\p}{\text{p}}

\newcommand{\GSB}{\text{GSB}}
\newcommand{\SE}{\text{SE}}
\newcommand{\ESA}{\text{ESA}}
\newcommand{\tr}{\text{tr}}
\newcommand{\E}{\hat{E}}
\newcommand{\C}{C}
\newcommand{\dt}{\Delta t}
\newcommand{\D}{D}

\newcommand{\nr}{\text{(nr)}}
\renewcommand{\r}{\text{(r)}}

\begin{document}
\title{Probing excited-state dynamics with quantum entangled photons: Correspondence to coherent multidimensional spectroscopy}

\author{Akihito Ishizaki}
\affiliation{Institute for Molecular Science, National Institutes of Natural Sciences, Okazaki 444-8585, Japan}\affiliation{School of Physical Sciences, Graduate University for Advanced Studies, Okazaki 444-8585, Japan}

\begin{abstract}
Quantum light is a key resource for promoting quantum technology. One such class of technology aims to improve the precision of optical measurements using engineered quantum states of light.
In this study, we investigate transmission measurement of frequency-entangled broadband photon pairs generated via parametric down-conversion with a monochromatic laser. It is observed that state-to-state dynamics in the system under study are temporally resolved by adjusting the path difference between the entangled twin beams when the entanglement time is sufficiently short. The non-classical photon correlation enables time-resolved spectroscopy with monochromatic pumping. It is further demonstrated that the signal corresponds to the spectral information along anti-diagonal lines of, for example, two-dimensional Fourier-transformed photon echo spectra. 
This correspondence inspires us to anticipate that more elaborately engineered photon states would broaden the availability of quantum light spectroscopy.
\end{abstract}
\date{\today}
\maketitle

\section{Introduction}

Ultrafast optical spectroscopy plays a pivotal role in investigating the structural and dynamic properties of complex molecules and materials. Most spectroscopic measurements project the microscopic information onto a single time or frequency axis, and hence, a wealth of information is difficult to extract unambiguously. To disentangle the information contents projected onto this one-dimensional axis, multidimensional observables need to be explored sometimes, in which the structural and dynamic information is projected onto more than two axes. Over the past quarter century, extensive effort has been devoted to the development of coherent multidimensional spectroscopy. \cite{Mukamel:2000io,SchlauCohen:2011cw,Fuller:2015kd,Kowalewski:2017jq}
To reveal ultrafast phenomena using spectroscopic methods, ultrashort pulsed lasers need to be applied. However, the broad bandwidth of such pulses prohibits the selective excitation of a single electronic state, making multidimensional spectra congested or even featureless. The key issue is that spectroscopy with classical light is subject to Fourier limitations on its joint temporal and spectral resolution. Further, this issue becomes more prominent when investigating biomolecular processes, such as photosynthetic light harvesting, in which multiple electronic states are present within a narrow energy range. 
As a possible solution, the polarization-specific technique has been employed in two-dimensional (2D) electronic and infrared spectroscopy. \cite{Hochstrasser:2001vp,Zanni:2001ks,Dreyer:2003th,SchlauCohen:2012dn,Westenhoff:2012fi}

On another front, quantum light, such as entangled photon pairs, is a key resource for promoting cutting-edge quantum technology. \cite{Walmsley:2015cn} One class of this technology aims to improve the precision of optical measurements via non-classical photon correlations. Quantum metrology has rapidly gained widespread attention due to its ability to make measurements with sensitivity and resolution beyond the limits imposed by the laws of classical physics. \cite{Simon2016book} In this light, it is hoped that quantum light will also open new avenues for optical spectroscopy using the parameters of quantum states of light. \cite{Dorfman:2016da,Schlawin:2018ci,Szoke:2020fc} Thus far, experiments of absorption spectroscopy with two-photon coincidence counting, \cite{Yabushita:2004hy,Kalachev:2013kh} two-photon absorption, \cite{Georgiades:1995dd,Dayan:2004kg,Dayan:2007bn,Lee:2006id,Harpham:2009cm,deJLeonMontiel:2019jt} two-photon induced fluorescence, \cite{Harpham:2009cm,Upton:2013is,Varnavski:2017eq} sum frequency generation, \cite{Peer:2005bc} and infrared spectroscopy with visible light \cite{Kalashnikov:2016cl} have been performed with frequency-entangled broadband photons generated through parametric down-conversion (PDC) or resonant hyper-parametric scattering. \cite{Inoue:2004ez,Edamatsu:2004ho}
In addition, selective two-photon excitation of a target state and control of the state distribution using the nonclassical photon correlation were theoretically investigated.\cite{Oka:2010if,Oka:2011dy,Schlawin:2012ba,Schlawin:2013dq,Lever:2019fd} A theoretical model for the scattering of an entangled photon pair from a molecular dimer were also developed.\cite{Bittner:2020gr}
Recently, special attention has been paid to the possibility of joint temporal and spectral resolutions. \cite{Fei:1997es,Saleh:1998vl,Dayan:2004kg,MacLean:2018by}
Entangled photon pairs are not subjected to the Fourier limitations on their joint temporal and spectral resolutions, \cite{Dayan:2004kg,Dayan:2007bn} and hence, the simultaneous improvement of time and frequency resolutions may be achievable. Motivated by this potential benefit, entangled photon-pair 2D fluorescence spectroscopy \cite{Raymer:2013kj,Dorfman:2014jm} and pump-probe and stimulated Raman spectroscopy with two-photon coincidence counting \cite{Schlawin:2016er,Dorfman:2014bn} were discussed.

In this work, we theoretically investigate the frequency-dispersed transmission measurement of frequency-entangled photon pairs generated via PDC pumped with a monochromatic laser.
In this spectroscopic method, the signal and idler photons are employed as the pump and probe fields, respectively, with delay interval. Then, we demonstrate that the non-classical correlation between the entangled photons enables time-resolved spectroscopy with monochromatic pumping instead of a pulsed laser. Moreover, the relation with heterodyned four-wave mixing measurement, such as 2D Fourier-transformed photon echo, and the influence of the entanglement time on the spectroscopic signals are described herein.

\section{Theory and results}

We consider electric fields inside a one-dimensional nonlinear crystal of length $L$ and subject to the PDC process. In this process, a pump photon with frequency $\omega_\p$ is split into signal and idler photons with frequencies $\omega_1$ and $\omega_2$, such that $\omega_\p = \omega_1 + \omega_2$. In the weak down-conversion regime, the state vector of the generated twin photons is written as \cite{Grice:1997ht,Keller:1997hj}
\begin{align}
	\lvert \psi_\pdc \rangle 
	\simeq
	\int d\omega_1 \int d\omega_2\,
	f(\omega_1, \omega_2) 
	\hat{a}_\s^\dagger(\omega_1) 
	\hat{a}_\i^\dagger(\omega_2) 
	\lvert \vac \rangle.
	\label{eq:state-vector}
\end{align}
In the equation $\lvert \vac \rangle $ denotes the photon vacuum state, and $\hat{a}_\s^\dagger(\omega)$ and $\hat{a}_\i^\dagger(\omega)$ are the creation operators of the signal and idler photons, respectively, of frequency $\omega$, where the commutation relation, $[\hat{a}_{\sigma}(\omega), \hat{a}_{\sigma'}^\dagger (\omega')]=\delta_{\sigma  \sigma'}\,\delta(\omega - \omega')$, is satisfied.
The two-photon amplitude, $f(\omega_1, \omega_2)$, is expressed as 
$f(\omega_1, \omega_2) = \zeta A_\p(\omega_1 + \omega_2)\,\phi (\omega_1 ,\omega_2 )$, where $A_\p (\omega)$ is the normalized pump envelope and 
$\phi (\omega_1,\omega_2) = \sinc[\Delta k(\omega_1, \omega_2)L/2]$ is the phase-matching function with momentum mismatch between the input and output photons, $\Delta k(\omega_1, \omega_2)$.
Typically, $\Delta k(\omega_1, \omega_2)$ may be approximated linearly around the central frequencies of the generated beams, $\bar\omega_\s$ and $\bar\omega_\i$, as
$\Delta k(\omega_1, \omega_2) L = (\omega_1 - \bar\omega_\s) T_\s + (\omega_2 - \bar\omega_\i) T_\i$ with $T_\sigma = L/v_\p - L/v_\sigma$, \cite{Rubin:1994ed,Keller:1997hj} where $v_\p$  and $v_{\sigma}$ are the group velocities of the pump laser and a generated beam at frequency $\bar\omega_\sigma$, respectively. The difference, $T_\en = \lvert T_\s-T_\i \rvert$, is referred to the entanglement time, \cite{Saleh:1998vl} which represents the maximum of the relative delay between the signal and idler photons. All other constants are merged into factor $\zeta$, which corresponds to the conversion efficiency of the PDC process.

We consider a system comprising molecules and light fields. The total Hamiltonian is written as  
\begin{align}
	\hat{H}_\text{total} 
	= 
	\hat{H}_\text{mol} + \hat{H}_\text{field} + \hat{H}_\text{mol--field}.
\end{align}
The first term, $\hat{H}_\text{mol}$, represents the Hamiltonian of photoactive degrees of freedom in the molecules, and the second term is the free Hamiltonian of the fields.
The electronic states are grouped into well-separated manifolds: electronic ground state $\lvert 0 \rangle$, single-excitation manifold $\{ \lvert e_\alpha \rangle \}$, and double-excitation manifold $\{ \lvert f_{\bar\gamma} \rangle \}$. 
In this paper, an overline such as $\bar\gamma$ indicates a state in the double-excitation manifold. 
Such level structures are typically found in molecular aggregates including photosynthetic pigment-protein complexes.\cite{Ishizaki:2012kf}
In this study direct two-photon absorption to the double-excitation manifold is not considered, and therefore, the optical transitions are described by the dipole operator, $\hat{\mu} = \hat{\mu}_+ + \hat{\mu}_-$, where $\hat{\mu}_+ $ is defined by 
\begin{align}
	\hat{\mu}_{+} 
	= 
	\sum_\alpha \mu_{\alpha 0} 
	\lvert e_\alpha  \rangle \langle 0 \rvert 
	+ 
	\sum_{\alpha \bar\gamma} \mu_{\bar\gamma \alpha} 
	\lvert f_{\bar\gamma} \rangle \langle e_\alpha \rvert,
	\label{eq:transitions}
\end{align}
and $\hat{\mu}_- = \hat{\mu}_+^\dagger$. Under the rotating-wave approximation, the molecule--field interaction can be written as 
$\hat{H}_\text{mol--field}(t) = -\hat{\mu}_+ \E^+(t) - \hat{\mu}_- \E^-(t)$,
where $\E^+(t)$ and $\E^-(t)$  denote the positive- and negative-frequency components, respectively, of the electric field operator.

\begin{figure}
	\includegraphics{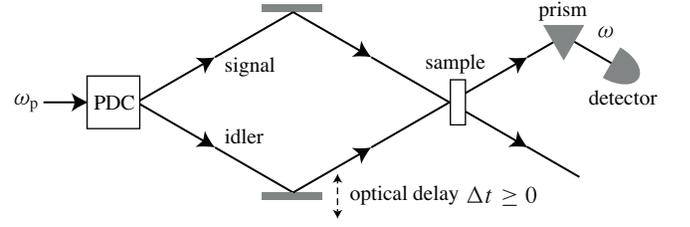}
	\caption{
	Schematic of frequency-dispersed transmission measurement with quantum entangled photon pairs generated via parametric down-conversion (PDC) pumped with a monochromatic laser of frequency $\omega_\p$. 
The signal--idler mutual delay interval is determined when generated at the PDC crystal. The upper bound of this innate interval is $T_\en/2$.
The delay interval is further controlled by adjusting the path difference between the signal and idler beams.
	The entangled photons are directed onto a sample with \textit{a posteriori} delay $\dt$ of the idler beam, and the change in the number of transmitted idler photons with frequency $\omega$ is registered.
	}
	\label{fig:1}
\end{figure}

The signal--idler relative delay is innately determined when generated; the upper bound of the interval is $T_\en/2$. However, the signal--idler delay interval is further controlled by adjusting the path difference between the beams. \cite{Hong:1987gm,Franson:1989go} This \textit{a posteriori} delay is denoted by $\dt$ herein.
Figure~\ref{fig:1} demonstrates the frequency-dispersed transmission measurement using the frequency-entangled photon pairs.
The same setup was discussed in Ref.~\onlinecite{schlawin:2013id}, as well as Refs.~\onlinecite{Roslyak:2009cy,Li:2017hc,Kalashnikov:2017hx}.
In this measurement, the signal photon is employed as the pump field, whereas the idler photon is used for the probe field with time delay $\dt \ge 0$, and hence, the positive-frequency component of the employed field operator is written as \cite{Hong:1987gm,Franson:1989go} 
\begin{align}
	\E^+(t) 
	= 
	\E_\i^+(t) + \E_\s^+(t+\dt), 
\end{align}
where $\E_\sigma^+(t) = (2\pi)^{-1}\int d\omega\, \hat{a}_\sigma(\omega)  e^{-i \omega t}$. The slowly varying envelope approximation has been adapted, with the bandwidth of the fields assumed to be negligible in comparison to the central frequency. \cite{loudon2000quantum} The probe field transmitted through sample $\E_\i$ is frequency-dispersed and the change in the transmitted photon number, $\E_\i^-(\omega) \E_\i^+(\omega)$, is measured. Thus, the frequency-dispersed intensity is written as \cite{Dorfman:2016da}
\begin{align}
	S(\omega; \dt)
	=
	\text{Im}
	\int^\infty_{-\infty} dt \, e^{i\omega t}
	\tr
	\left[ \E_\i^-(\omega) \hat\mu_- \hat\rho(t) \right],
	\label{eq:transmission}
\end{align}
with the initial condition of $\hat\rho(-\infty) = \lvert 0 \rangle\langle 0 \rvert \otimes \lvert \psi_\pdc \rangle\langle \psi_\pdc \rvert$. The lowest-order contribution of Eq.~\eqref{eq:transmission} is the absorption of only the idler photon. However, the absorption signal is independent of the PDC pump frequency, $\omega_\p$, and delay time, $\dt$. Thus, the process can be in principle separated from the pump-probe-type two-photon process, although the separation is experimentally difficult due to much smaller nonlinear contribution. 
Consequently, the perturbative expansion of $\hat\rho(t)$ with respect to the molecule--field interaction, $\hat{H}_\text{mol--field}$, yields the third-order term as the leading order contribution.
The resultant signal is expressed as the sum of eight contributions, which are classified into ground-state bleaching (GSB), stimulated emission (SE), excited-state absorption (ESA), and double-quantum coherence (DQC). Typically, the coherence between the electronic ground state and a doubly excited state rapidly decays in comparison to the others, and hence, the DQC contribution is disregarded in this work.  
Each contribution is written as
\begin{multline}
	S_x^{(y)}(\omega; \dt)
	=
	\mathrm{Im}
	\int^\infty_{-\infty}dt\,e^{i\omega t}
\\\times
	\iiint^\infty_0 d^3{\bf s}\,
	\Phi_x^{(y)}(s_3, s_2, s_1)
	\C_x^{(y)}(\omega, t; s_3, s_2, s_1),
	\label{eq:each-signal}
\end{multline}
where $x$ indicates GSB, SE, or ESA, and $y$ indicates ``rephasing'' (r) or ``non-rephasing'' (nr). The function of $\Phi_x^{(y)}(s_3,s_2,s_1)$ indicates the third-order response function of the molecules, whereas $\C_x^{(y)}(\omega, t; s_3, s_2, s_1)$ is the four-body correlation function of the field operators, such as
$	\C_\ESA^\r(\omega, t; s_3, s_2, s_1) 
	=
	\langle
		\E^-(t-s_3-s_2-s_1)
		\E^-_\i(\omega)
		\E^+(t-s_3)
		\E^+(t-s_3-s_2)
	\rangle$.
The bracket indicates the expectation value in terms of $\lvert \psi_\pdc \rangle$, namely $\langle \dots \rangle = \langle \psi_\pdc \vert \dots \vert \psi_\pdc \rangle$.

To obtain a concrete but simple expression of the signal, the memory effect straddling different time intervals in the response function is ignored. \cite{Ishizaki:2008be} Consequently, the response function is expressed in a simpler form,
$
	\Phi(t_3,t_2,t_1)
	=
	(i/\hbar)^3 
	\tr[
		\hat\mu_-
		\hat{G}(t_3)
		\hat\mu^\times
		\hat{G}(t_2)
		\hat\mu^\times
		\hat{G}(t_1)
		\hat\mu^\times
		\lvert 0 \rangle\langle 0 \rvert
	]
$,
where the trace is computed only for the photoactive degrees of freedom, $\lvert 0 \rangle$, $\{\lvert e_\alpha \rangle\}$, and $\{\lvert f_{\bar\gamma}\rangle\}$. 
In the equation, $\hat{G}(t)$ denotes the time-evolution operator of the molecular excitations and the super-operator notation was introduced, $\hat{\mu}^\times\hat{O} = [\hat{\mu},\hat{O}]$ for any operand $\hat{O}$. Hereafter, the reduced Planck constant $\hbar$ will be omitted. For example, the rephasing contribution of the ESA signal is written as\cite{Ishizaki:2012kf}
\begin{multline}
	\Phi_\ESA^\r(t_3,t_2, t_1) 
	= 
	-i^3 
	\sum_{\alpha\beta\gamma\delta\bar\epsilon} 
	\mu_{\delta\bar \epsilon}
	\mu_{\bar\epsilon \gamma}
	\mu_{\alpha 0}
	\mu_{0 \beta}
\\
	\times
	G_{\bar\epsilon\delta}(t_3) 
	G_{\gamma\delta \gets \alpha\beta}(t_2) 
	G_{0\beta}(t_1),
\end{multline}
where $G_{\gamma\delta \gets \alpha\beta}(t)$ is the matrix element of the time-evolution operator defined by $\rho_{\gamma\delta}(t) = \sum_{\alpha\beta} G_{\gamma\delta \gets \alpha\beta}(t-s) \rho_{\alpha\beta}(s)$, and $G_{\alpha\beta}(t)$ describes the time evolution of the $\lvert e_\alpha \rangle\langle e_\beta \rvert$ coherence.

To calculate the signal, the four-body correlation functions of the field operators also need to be computed. To this end, we consider a case of monochromatic pumping with frequency $\omega_\p$ for the PDC process. The two-photon amplitude in Eq.~\eqref{eq:state-vector} is recast into
$
	f(\omega_1,\omega_2 ) 
	= 
	\zeta \delta(\omega_1+\omega_2-\omega_\p)\, 
	\sinc[(\omega_2 - \bar\omega_\i)T_\en/2]
$.
Consequently, the two-photon wave function and field auto-correlation function, which appear in normal ordering in the four-body correlation functions, are computed as
\begin{gather}
	\langle \vac \vert
		\E_\s^+(t) \E_\i^+(s)
	\vert \psi_\pdc \rangle
	=
	\frac{\zeta }{2\pi} 
	\D_1(t-s)
	e^{-i\bar\omega_\s t-i\bar\omega_\i s},
	\label{eq:2photon-wavefunc}
\\
	\langle \psi_\pdc \vert \E_\sigma^-(t) \E_\sigma^+(s) \vert \psi_\pdc \rangle
	=
	\frac{\zeta^2}{2\pi}
	\D_2(t-s)
	e^{i\bar\omega_\sigma (t-s)},
	\label{eq:correlation-func}
\end{gather}
where $\D_n(t)$ is defined by
\begin{align}
	\D_n(t)
	=
	\int^\infty_{-\infty}
	\frac{d\omega}{2\pi}
	e^{-i\omega t} 
	\left(
		\sinc\frac{\omega T_\en}{2} 
	\right)^n.
\end{align}
The expressions of $\D_1(t)$ and $\D_2(t)$ are calculated by the rectangular and triangular functions of $t/T_\en$, respectively. In this work, however, our attention is directed to the limit of $T_\en\to 0$,
\begin{align}
	\lim_{T_\en\to 0} \D_n(t) 
	= 
	\delta(t),
	\label{eq:small-TE}
\end{align}
which holds true irrespective of the values of $n$. 
This implies that the four-body correlation functions can be simply expressed when the entanglement time, $T_\en$, is sufficiently short as compared to the characteristic timescales of the dynamics under investigation. \cite{Fujihashi:2020ep}
Indeed, the four-body correlation function in the rephasing ESA signal reduces to
\begin{multline}
	\C_\ESA^\r(\omega, t; s_3, s_2, s_1) 
\\
	=
	\delta(s_2-\dt)
	e^{	-i\omega t +i\omega s_3 -i(\omega_{\rm p} -\omega)s_1}.
\end{multline}
The case of longer entanglement time will be discussed later in this paper. 

Provided that electronic coherence in the single-excitation manifold rapidly decays and thus negligible, Eq.~\eqref{eq:transmission} is expressed as
\begin{multline}
	S(\omega;\dt)
	=
	S_\GSB(\omega; \dt)
	+
	S_\SE(\omega; \dt)
\\
	+
	S_\ESA(\omega; \dt)
	+
	S_\text{c}(\omega),
	\label{eq:transmission-1}
\end{multline}
in terms of the GSB, SE, and ESA contributions,
\begin{align}
	S_\GSB(\omega; \dt)
	&=
	-
	\sum_{\alpha\beta} 
	I_{\beta 0, \alpha 0}(\omega,\omega_\p-\omega)
	G_{00\gets 00}(\dt), 
	\label{eq:GSB}
\\
	S_\SE(\omega; \dt)
	&\simeq
	-
	\sum_{\alpha\beta} 
	I_{\beta 0, \alpha 0}(\omega,\omega_\p-\omega)
	G_{\beta\beta\gets \alpha\alpha}(\dt),
	\label{eq:SE}
\\
	S_\ESA(\omega; \dt)
	&\simeq
	+
	\sum_{\alpha\beta\bar\epsilon} 
	I_{\bar\epsilon \beta,\alpha 0}(\omega,\omega_\p-\omega)	
	G_{\beta\beta\gets \alpha\alpha}(\dt).
	\label{eq:ESA}
\end{align}
In the above, $I_{\alpha\beta,\gamma\delta}(\omega_2,\omega_1)$ is defined by 
\begin{align}
	I_{\alpha\beta,\gamma\delta}(\omega_2,\omega_1) 
	= 
	\mu_{\alpha \beta}^2 
	\mu_{\gamma\delta}^2
	G'_{\alpha \beta}[\omega_2]
	G'_{\gamma\delta}[\omega_1],
	\label{eq:absorptive-2D}
\end{align}
where $G'_{\alpha\beta}[\omega]$ is the real part of the Fourier--Laplace transform of the time-evolution operator, $G_{\alpha\beta}[\omega] = \int^\infty_0 dt\, e^{i\omega t} G_{\alpha\beta}(t)$. 
The last term in Eq.~\eqref{eq:transmission-1} originates from the auto-correlation function in Eq.~\eqref{eq:correlation-func}, and is written as
$
	S_\text{c}(\omega)
	=
	-
	\sum_{\alpha\beta} 
	\mu_{\beta 0}^2 \mu_{\alpha 0}^2
	G'_{\beta 0}[\omega]
	( G'_{\alpha 0}[\omega] + K_{\beta\alpha} )
$
with $K_{\beta\alpha} = \int^\infty_0 ds\, G_{\beta\beta \gets \alpha\alpha}(s)$.
It is noted the last term in Eq.~\eqref{eq:transmission-1} are independent of $\dt$.\footnote{In Ref.~\onlinecite{Schlawin:2017ea}, it was discussed that the field commutator, which appear in normal ordering in the four-body correlation functions,  gives rise to the exchange of virtual photons between transition dipoles and thus dipolar interactions in molecular aggregates. In this work, however, we only focus on the fact that the last term in Eq.~\eqref{eq:transmission-1} are independent of $\dt$ for simplicity.}
In deriving Eq.~\eqref{eq:transmission-1}, we employed the approximation of $G_{\beta\beta \gets \alpha\alpha}(\dt-s_1) G_{\alpha 0}(s_1) \simeq G_{\beta\beta \gets \alpha\alpha}(\dt) G_{\alpha 0}(s_1)$ for the non-rephasing Liouville pathways. \cite{Cervetto:2004gm}
This approximation is justified when the response function varies slowly as a function of the waiting time, $t_2$. Namely, the dynamics within the single-excitation manifold are slow in comparison to the decay of the coherence between different manifolds during the $t_1$ period.
To remove the $\dt$-independent contributions, the difference spectrum is considered,
$
	\Delta S(\omega; \dt) 
	= 
	S(\omega; \dt) - S(\omega; \dt = 0)
$,
which contains only the SE and ESA contributions as a function of $\dt$.
When the electronic coherence in the single-excitation manifold is considered, the following terms need to be added to Eqs.~\eqref{eq:SE} and \eqref{eq:ESA}:
\begin{multline}
	S_\SE^\text{(coh)}(\omega; \dt)
	\simeq
	-\text{Re}
	\sum_{\alpha\beta\gamma\delta} 
	\mu_{\delta 0}
	\mu_{\gamma 0}
	\mu_{\beta  0} 
	\mu_{\alpha 0}
	G_{\gamma 0}[\omega]
\\	
	\times
	\left(
		G_{\alpha 0}[\omega_\p-\omega] + G_{\beta  0}^\ast[\omega_\p-\omega]
	\right)
	G_{\gamma\delta \gets \alpha\beta}(\dt)
	\label{eq:SE-coh}
\end{multline}
and
\begin{multline}
	S_\ESA^\text{(coh)}(\omega; \dt)
	\simeq
	+\text{Re}
	\sum_{\alpha\beta\gamma\delta\bar\epsilon} 
	\mu_{\bar\epsilon \delta }
	\mu_{\bar\epsilon \gamma}
	\mu_{\beta  0}
	\mu_{\alpha 0}
	G_{\bar\epsilon \delta}[\omega] 
\\\times
	\left(
		G_{\alpha 0}[\omega_\p-\omega] + G_{\beta 0}^\ast[\omega_\p-\omega]
	\right)
	G_{\gamma\delta \gets \alpha\beta}(\dt),
	\label{eq:ESA-coh}
\end{multline}
where the bath-induced coherence transfer,\cite{Ohtsuki:1989fv,Jean:1995ke} $\lvert \alpha\rangle\langle \beta \rvert \to \lvert \gamma\rangle\langle \delta\rvert$ ($\alpha\ne\gamma$ and/or $\beta\ne\delta$), were also included.
Notably, the non-classical correlation between the twin photons enables time-resolved spectroscopy with monochromatic pumping.

To obtain the information contents of the signal, we assume that the time evolution in the $t_1$ and $t_3$ periods is described as $G_{\alpha\beta}(t) = e^{-(i\omega_{\alpha\beta}+\epsilon_+)t}$, thereby leading to the expression of $I_{\beta 0, \alpha 0}(\omega,\omega_\p)$ in the SE contribution,
$I_{\beta 0, \alpha 0}(\omega,\omega_\p-\omega)	\propto \delta(\omega - \omega_{\beta 0}) \delta(\omega_\p - \omega - \omega_{\alpha 0})$.
It can be understood that the stimulated emission probed at frequency $\omega = \omega_{\beta 0}$ indicates that an excited state of electronic energy $\omega_{\alpha0} = \omega_\p - \omega_{\beta 0}$ was populated with the pump field.
Similarly, the ESA contribution with frequency $\omega = \omega_{\bar\gamma \beta}$ is also understood.
The non-classical correlation between the entangled twin photons restricts possible optical transitions for a given PDC pump frequency.
Therefore, Eq.~\eqref{eq:transmission-1} spectrally resolves a pair of optical transitions, $(0 \to e_\alpha,0 \to e_\beta)$ or $(0 \to e_\alpha, e_\beta \to f_{\bar\gamma})$, with the pump frequency $\omega_\p$ for PDC, provided that the equality of $\omega_\p = \omega_{\alpha 0} + \omega_{\beta 0}$ or $\omega_\p = \omega_{\alpha 0} +\omega_{\bar\gamma \beta}$ holds.
Simultaneously, Eq.~\eqref{eq:transmission-1} temporally resolves the excited state dynamics of $e_\alpha \to e_\beta$ through the intensity change of the SE signal at frequency $\omega = \omega_{\beta 0}$ or the ESA signal at frequency $\omega = \omega_{\bar\gamma \beta}$.

\section{Discussion}

It is noted that the above property corresponds to the anti-diagonal cut of a coherent 2D optical spectrum. Thus, we consider the absorptive 2D spectrum obtained with a pump-probe \cite{Cervetto:2004gm} or photon-echo technique \cite{Khalil:2003fn} in the impulsive limit. 
The absorptive 2D spectrum is obtained from the sum of the rephasing and non-rephasing contributions and expressed as 
\begin{multline}
	\mathcal{S}_{\rm 2D}(\omega_3,t_2,\omega_1)
	=
	\mathcal{S}_\GSB(\omega_3,t_2,\omega_1)
\\
	+
	\mathcal{S}_\SE (\omega_3,t_2,\omega_1)
	+
	\mathcal{S}_\ESA(\omega_3,t_2,\omega_1),
\end{multline}
with the  GSB, SE, and ESA contributions
\begin{align}
	\mathcal{S}_\GSB(\omega_3,t_2,\omega_1)
	&=
	+
	\sum_{\alpha\beta} 
	I_{\beta 0, \alpha 0}(\omega_3,\omega_1)
	G_{00 \gets 00}(t_2),
	\label{eq:GSB-2D}
\\
	\mathcal{S}_\SE(\omega_3,t_2,\omega_1)
	&=
	+
	\sum_{\alpha\beta} 
	I_{\beta 0, \alpha 0}(\omega_3,\omega_1)
	G_{\beta\beta \gets \alpha\alpha}(t_2),
	\label{eq:SE-2D}
\\
	\mathcal{S}_\ESA(\omega_3,t_2,\omega_1)
	&=
	-
	\sum_{\alpha\beta\bar\gamma} 
	I_{\bar\gamma \beta,\alpha 0}(\omega_3,\omega_1)
	G_{\beta\beta \gets \alpha\alpha}(t_2).
	\label{eq:ESA-2D}
\end{align}
Notably, Eq~\eqref{eq:transmission-1} provides the spectral information along the anti-diagonal line, $\omega_1 + \omega_3 = \omega_\p$, on the absorptive 2D spectrum,
\begin{align}
	S(\omega_3;\dt) 
	= 
	- \mathcal{S}_{\rm 2D}(\omega_3, \dt, \omega_\p - \omega_3), 
	\label{eq:correspondence}
\end{align}
except for the $\dt$-independent term in Eq.~\eqref{eq:transmission-1}, $S_\text{c}(\omega)$. 
The equality still holds true when the coherence contributions in Eqs.~\eqref{eq:SE-coh} and \eqref{eq:ESA-coh} are considered.
Therefore, the appropriate selection of pump frequency $\omega_\p$ allows one to analyze individual diagonal and/or off-diagonal peaks of the 2D spectrum.
Furthermore, by sweeping pump frequency $\omega_\p$, the transmission intensity, $S(\omega;\dt)$, becomes homologous to the 2D spectrum, $\mathcal{S}_{\rm 2D}(\omega_3, \dt,\omega_1)$.

However, it should not be overlooked that the correspondence between $S(\omega;\dt)$ and  $\mathcal{S}_{\rm 2D}(\omega_3,t_2,\omega_1)$ is only true for the condition of short entanglement time. To demonstrate the difference in the case of longer entanglement time, the rephasing contribution of the ESA signal is considered as an example. For a finite value of the entanglement time, the four-body correlation function of the field is computed as
\begin{multline}
	\C_\ESA^\r(\omega,t;s_3,s_2,s_1)
\\
	=
	{\rm sinc}\frac{(\omega-\bar\omega_\i)T_\en}{2}
	e^{	-i\omega t +i\omega s_3 -i(\omega_{\rm p}-\omega)s_1 -i \Omega_\i \dt}
\\ \times
	[ \D_1(s_2-\dt) e^{i\Omega_\i s_2} + \D_1(s_2+\dt) e^{i\Omega_\s s_2} ],
	\label{eq:4body-r-ESA}
\end{multline}
where $\D_1(t) = T_\en^{-1}\mathrm{rect}(t/T_\en)$ and $\Omega_\sigma = \omega - \bar\omega_\sigma$.
Thus, the rephasing contribution of the ESA signal is obtained as
\begin{multline}
	S_\ESA^\r(\omega;\dt)
	=
	\sinc\frac{(\omega-\bar\omega_\i)T_\en}{2}
	\text{Re}
	\sum_{\alpha\beta\gamma\delta\bar\epsilon} 
	\mu_{\bar\epsilon \delta}
	\mu_{\bar\epsilon \gamma}
	\mu_{\beta  0}
	\mu_{\alpha 0}
\\\times
	G_{\bar\epsilon \delta}[\omega] 
	G_{\beta 0}^\ast[\omega_\p-\omega]
	F_{\gamma\delta\gets\alpha\beta}(\omega,\dt),
	\label{eq:signal-r-ESA}
\end{multline}
where 
$F_{\gamma\delta\gets\alpha\beta}(\omega,\dt)$ is introduced as
\begin{multline}
	F_{\gamma\delta\gets\alpha\beta}(\omega,\dt)
	=
	\int^\infty_0 ds \,
	G_{\gamma\delta \gets \alpha\beta}(s) 
	e^{-i \Omega_\i \dt}
\\\times
	[ \D_1(s-\dt) e^{i\Omega_\i s} + \D_1(s+\dt) e^{i\Omega_\s s} ].
	\label{eq:F-expression}
\end{multline}
The rephasing SE signal is also expressed by a similar formula including $F_{\gamma\delta \gets \alpha\beta}(\omega,\dt)$. 
However, the non-rephasing SE and ESA signals are expressed with more complicated equations owing to $\D_1(s_2+s_1\pm \dt)$.
When the coherences between the electronic eigenstates in the single-excitation manifold are ignored and energy transfer rates between the electronic eigenstates are provided, the matrix elements of the time-evolution operator, 
$G_{\gamma\delta \gets \alpha\beta}(t)$, 
are written as the sum of exponential functions of $t$.
For demonstration purposes, we model the matrix element as 
$G_{\gamma\delta \gets \alpha\beta}(t) = e^{-\lambda t}$. 
Here, it should be noted that the interval between the arrival times of the signal and idler photons at the molecular sample, $t_\i - t_\s$, becomes blurred because of the entanglement time, $T_\en$, as
\begin{align}
	{\dt - \frac{T_\en}{2}}  \le  {t_\i - t_\s}  \le  {\dt + \frac{T_\en}{2}}.
\end{align}
However, inequality $t_\i - t_\s \ge 0$ should hold for the pump-probe-type two-photon process depicted in Fig.~\ref{fig:1}.
In the case of $T_\en/2 < \dt$, the expression of 
$F_{\gamma\delta \gets \alpha\beta}(\omega,\dt)$
is obtained as
\begin{align}
	F_{\gamma\delta\gets\alpha\beta}(\omega,\dt)
	&=
	\sinc\frac{(\omega-\bar\omega_\i+i\lambda)T_\en}{2}
	e^{-\lambda\dt}.
	\label{eq:t-finite-te:1}
\end{align}
In contrast, in the case of $\dt < T_\en/2$,  Eq.~\eqref{eq:F-expression} leads to
\begin{multline}
	F_{\gamma\delta\gets\alpha\beta}(\omega,\dt)
	=
	\frac{
		e^{i(\omega - \bar\omega_\i + i\lambda)T_\en/2} 
		e^{-\lambda \dt} 
		- 
		e^{-i(\omega - \bar\omega_\i)\dt}
	}{	
		i (\omega - \bar\omega_\i + i\lambda)T_\en
	}
\\
	+
	\frac{
		e^{i(\omega - \bar\omega_\s + i\lambda)T_\en/2}
		e^{-i(2\omega - \omega_\p + i\lambda)\dt} 
		- 
		e^{-i(\omega - \bar\omega_\i)\dt}
	}{
		i (\omega-\bar\omega_\s + i\lambda)T_\en
	}.
	\label{eq:t-finite-te:2}
\end{multline} 
If coherence $\lvert\alpha\rangle\langle\beta\rvert$ is considered, the time-evolution operator is modeled as  
$G_{\gamma\delta \gets \alpha\beta}(t) = e^{-i\omega_{\alpha\beta} t - \lambda t}$, 
and $\lambda$ in Eqs.~\eqref{eq:t-finite-te:1} and \eqref{eq:t-finite-te:2} is replaced with $i\omega_{\alpha\beta}+\lambda$.
When the entanglement time is sufficiently short in comparison to the characteristic timescale of dynamics under investigation, Eq.~\eqref{eq:t-finite-te:1} leads to 
$
	F_{\gamma\delta \gets \alpha\beta}(\omega,\dt) \simeq G_{\gamma\delta \gets \alpha\beta}(\dt)$.
However, in the opposite limit, Eq.~\eqref{eq:t-finite-te:2} exhibits complicated time evolution depending on the values of $\bar\omega_\s$ and $\bar\omega_\i$, and hence, it is impossible to extract relevant information on the excited-state dynamics from the signal. 
Moreover, 
Eqs.~\eqref{eq:signal-r-ESA} -- \eqref{eq:t-finite-te:2} 
demonstrate that the signal as a function of entanglement time, $T_\en$, does not provide direct information on the excited-state dynamics such as $G_{\gamma\delta \gets \alpha\beta}(T_\en)$, when the electronic transitions in Eq.~\eqref{eq:transitions} and the optical system depicted in Fig.~\ref{fig:1} are considered.

\section{Concluding remarks}

In this work, we theoretically investigated quantum entangled two-photon spectroscopy, specifically frequency-dispersed transmission measurement using frequency-entangled photon pairs generated via PDC pumped with a monochromatic laser. When the entanglement time is sufficiently short compared to characteristic timescales of the dynamics under investigation, the transmission measurement is capable of temporally resolving the state-to-state dynamics, although a monochromatic laser is employed. 
Furthermore, we demonstrated that the transmission measurement could provide the same information contents as in heterodyned four-wave mixing signals, such as 2D Fourier-transformed photon echo, although a simple optical system and simple light source are employed. This correspondence inspires us to anticipate that the usage of more elaborately engineered quantum states of light \cite{Dinani:2016gd,Cho:2018cf,Ye:2020dz} would broaden the availability of quantum light spectroscopy \cite{Mukamel:2020ej} or molecular quantum metrology. The extensions of the present work in these directions are to be explored in future studies.

\begin{acknowledgements}
The author is grateful to Yuta Fujihashi for his assistance in preparing the manuscript. He would also like to thank Animesh Datta and Yutaka Shikano for their valuable comments on the manuscript. This work was supported by JSPS KAKENHI Grant No.~17H02946, MEXT KAKENHI Grant No.~17H06437 in Innovative Areas ``Innovations for Light--Energy Conversion,'' and MEXT Quantum Leap Flagship Program Grant No.~JPMXS0118069242.
\end{acknowledgements}

\appendix
\section{Four-body correlation functions of the electric field operators}

To calculate the signal, the four-body correlation functions of the electric field operator, $\E(t) = \E_\i(t) + \E_\s(t+\dt)$, need to be computed.
With the use of Eqs.~\eqref{eq:2photon-wavefunc} and \eqref{eq:correlation-func}, the four-body correlation functions in Eq.~\eqref{eq:each-signal} are computed as follows:
\begin{align}
	&\C_\GSB^\r(\omega,t;s_3,s_2,s_1)
\notag\\
	&=
	\langle
		\hat{E}^-(t-s_3-s_2-s_1)
		\hat{E}^+(t-s_3-s_2)
		\hat{E}^-_\i(\omega)
		\hat{E}^+(t-s_3)
	\rangle
\notag\\
	&=	
	\tilde\D(\Omega_\i)
	\D_1(s_2-\dt)
	e^{	-i\omega t
		+i\omega s_3
		+i\Omega_\i s_2
		-i(\omega_{\rm p}-\omega)s_1
		-i\Omega_\i\dt}
\notag\\
	&
	+
	\tilde\D(\Omega_\i)
	\D_1(s_2+\dt)
	e^{-i\omega t
		+i\omega s_3
		+i\Omega_\s s_2
		-i(\omega_{\rm p}-\omega)s_1
		-i\Omega_\i\dt} 
\notag\\
	&
	+
	\D_2(s_2+s_1)
	e^{	-i\omega t
		+i\omega s_3
		+i\Omega_\s s_2
		-i\bar\omega_\s s_1} 
	,
\end{align}
\begin{align}
	&\C_\GSB^\nr(\omega,t;s_3,s_2,s_1)
\notag\\
	&=
	\langle 
		\hat{E}^-_\i(\omega)
		\hat{E}^+(t-s_3)
		\hat{E}^-(t-s_3-s_2)
		\hat{E}^+(t-s_3-s_2-s_1)
	\rangle
\notag\\
	&=
	\tilde\D(\Omega_\i)
	\D_1(s_2+s_1-\dt)
	e^{	-i\omega t
		+i\omega s_3
		+i\Omega_\i s_2
		+i\bar\omega_\s s_1
		-i\Omega_\i \dt
	}
\notag\\
	&
	+ 
	\tilde\D(\Omega_\i)
	\D_1(s_2+s_1+\dt)
	e^{	-i\omega t
		+i\omega s_3
		+i\Omega_\s s_2
		+i\bar\omega_\i s_1
		-i\Omega_\i \dt
	}
\notag\\
	&
	+
	\tilde\D(\Omega_\i)^2
	\delta(s_2)
	e^{	-i\omega t
		+i\omega s_3
		+i\omega s_1}
	,
\end{align}
\begin{align}
	&\C_\SE^\r(\omega,t;s_3,s_2,s_1)
\notag\\
	&=
	\langle
		\hat{E}^-(t-s_3-s_2-s_1)
		\hat{E}^+(t-s_3)
		\hat{E}^-_\i(\omega)
		\hat{E}^+(t-s_3-s_2)
	\rangle
\notag\\
	&
	=
	\tilde\D(\Omega_\i)
	\D_1(s_2-\dt)
	e^{	-i\omega t
		+i\omega s_3
		+i\Omega_\i s_2
		-i(\omega_{\rm p}-\omega)s_1
		-i\Omega_\i\dt}
\notag\\
	&
	+
	\tilde\D(\Omega_\i)
	\D_1(s_2+\dt)
	e^{	-i\omega t
		+i\omega s_3
		+i\Omega_\s s_2
		-i(\omega_{\rm p}-\omega)s_1
		-i\Omega_\i\dt}
\notag\\
	&
	+
	\D_2(s_1)
	e^{-i\omega t
	+i\omega s_3
	-i\bar\omega_\s s_1} 
	,
\end{align}
\begin{align}
	&\C_\SE^\nr(\omega,t;s_3,s_2,s_1)
\notag\\
	&=
	\langle 
		\hat{E}^-(t-s_3-s_2)
		\hat{E}^+(t-s_3)
		\hat{E}^-_\i(\omega)
		\hat{E}^+(t-s_3-s_2-s_1)
	\rangle	
\notag\\
	&
	=
	\tilde\D(\Omega_\i)
	\D_1(s_2+s_1-\dt)
	e^{	-i\omega t
		+i\omega s_3
		+i\Omega_\i s_2
		+i\bar\omega_\s s_1
		-i\Omega_\i \dt}
\notag\\
	&+
	\tilde\D(\Omega_\i)
	\D_1(s_2+s_1+\dt)
	e^{	-i\omega t
		+i\omega s_3
		+i\Omega_\s s_2
		+i\bar\omega_\i s_1
		-i\Omega_\i\dt}
\notag\\
	&
	+
	\D_2(s_1)
	e^{	-i\omega t
		+i\omega s_3
		i\bar\omega_\s s_1},
\end{align}
\begin{align}
	&\C_\ESA^\r(\omega,t;s_3,s_2,s_1)
\notag\\
	&=
	\langle
		\hat{E}^-(t-s_3-s_2-s_1)
		\hat{E}^-_\i(\omega)
		\hat{E}^+(t-s_3)
		\hat{E}^+(t-s_3-s_2)
	\rangle	
\notag\\
	&=
	\tilde\D(\Omega_\i)
	\D_1(s_2-\dt)
	e^{	-i\omega t
		+i\omega s_3
		+i\Omega_\i s_2
		-i(\omega_{\rm p}-\omega)s_1
		-i\Omega_\i\dt}
\notag\\
	&
	+
	\tilde\D(\Omega_\i)
	\D_1(s_2+\dt)
	e^{	-i\omega t
		+i\omega s_3
		+i\Omega_\s s_2
		-i(\omega_{\rm p}-\omega) s_1
		-i\Omega_\i \dt},
\end{align}
and
\begin{align}
	&\C_\ESA^\nr(\omega,t;s_3,s_2,s_1)
\notag\\
	&=
	\langle
		\hat{E}^-(t-s_3-s_2)
		\hat{E}^+(t-s_3)
		\hat{E}^-_\i(\omega)
		\hat{E}^+(t-s_3-s_2-s_1)
	\rangle
\notag\\
	&
	=
	\tilde\D(\Omega_\i)
	\D_1(s_2+s_1-\dt)
	e^{	-i\omega t
		+i\omega s_3	
		+i\Omega_\i s_2
		+i\bar\omega_\s s_1
		-i\Omega_\i \dt}
\notag\\
	&
	+
	\tilde\D(\Omega_\i)
	\D_1(s_2+s_1+\dt)
	e^{	-i\omega t
		+i\omega s_3
		+i\Omega_\s s_2
		+i\bar\omega_\i s_1
		-i\Omega_\i \dt},
\end{align}
where $\tilde{\D}(\omega) = {\rm sinc}({\omega T_\en}/{2})$ is defined, and the common prefactor of each term, ${\zeta^2}/{(2\pi)^2}$, is omitted. 
In the limit of $T_\en \to 0$, we obtain $\tilde\D(\omega) \simeq 1$ and $\D_1(t) \simeq \delta(t)$.


%

\end{document}